\documentclass[twocolumn,amsmath,amssymb,pra,superscriptaddress]{revtex4-1}

\usepackage{bm}

\usepackage[final,letterspace=-40]{microtype}
\usepackage[colorlinks=true, linkcolor=blue, urlcolor=blue,  citecolor=blue, anchorcolor=blue]{hyperref}

\usepackage{mathpazo} 

\usepackage{graphicx}
\usepackage{dcolumn}
\usepackage{bm}
\usepackage{helvet}

\usepackage{caption}
\DeclareCaptionLabelFormat{plain}{Figure #2 $\boldsymbol |$ }

\makeatletter
\renewcommand*{\@fnsymbol}[1]{\ensuremath{\ifcase#1\or *\or *,\dagger\or \ddagger\or
    \mathsection\or \mathparagraph\or \|\or **\or \dagger\dagger
    \or \ddagger\ddagger \else\@ctrerr\fi}}
\makeatother

\begin{document}

\title{Compressive optical interferometry}
\author{Davood Mardani}
\thanks{These authors contributed equally.}
\affiliation{Dept. of Electrical and Computer Engineering, University of Central Florida, Orlando, FL 32816, USA}
\author{H. Esat Kondakci} 
\thanks{These authors contributed equally.}
\author{Lane Martin}
\author{Ayman F. Abouraddy}
\affiliation{CREOL, The College of Optics \& Photonics, University of Central Florida, Orlando, Florida 32816, USA}
\author{George K. Atia}
\affiliation{Dept. of Electrical and Computer Engineering, University of Central Florida, Orlando, FL 32816, USA}

\begin{abstract} \noindent 	Compressive sensing (CS) combines data acquisition with compression coding to reduce the number of measurements required to reconstruct a sparse signal. In optics, this usually takes the form of projecting the field onto sequences of random spatial patterns that are selected from an appropriate random ensemble. We show here that CS can be exploited in `native' optics hardware without introducing added components. Specifically, we show that random sub-Nyquist sampling of an interferogram helps reconstruct the field modal structure. The distribution of reduced sensing matrices corresponding to random measurements is provably incoherent and isotropic, which helps us carry out CS successfully.
\end{abstract}

\maketitle


\noindent 
Compressive sensing (CS) is a strategy for reducing the number of measurements required to reconstruct a signal by projecting it onto a basis of random functions, thereby combining the two steps of \textit{data acquisition} and \textit{compression coding} \cite{Donoho06TIT,Candes08CR}. This procedure is particularly effective when the signal is \textit{sparse} -- i.e., the signal receives contributions from a small number of basis functions in some representation \cite{Donoho06TIT}. Underlying this approach is an under-determined linear transformation between the sparse representation and the measurement basis called a \textit{sensing matrix} that is amenable to a stable pseudoinversion \cite{Candes06PICM, Donoho06TIT}. CS has found many uses in optics such as the `single-pixel' camera \cite{Duarte08SPM, Magalhaes11AO}, holography \cite{Martinez-Leon17, Clemente13, Deepan14SPIE, Yair13OL}, optical imaging and tomography \cite{Liutkus14SR, Graff15AO, Mehmet10OE,Yao15BOE}, optical encryption \cite{Li15SCIrep, Rawat15AP}, spectral analysis \cite{Zhu14OE}, and efficient quantum state tomography \cite{Howland14PRL, Mirhosseini14PRL, Kalev15QI, Howland16PRX}. In all these cases, the native optical system is modified by incorporating a new component that projects the field onto \textit{random} patterns. CS then reduces the number of data samples acquired by the \textit{modified} system that are needed to recover the target sparse vector. 

In this paper, we develop a compressive approach to interferometry for optical modal analysis in arbitrary bases and realize it experimentally. We underscore four major differences from related work. First, unlike much of the prior work which introduces random masks along the optical path to obtain randomized measurements \cite{Howland16PRX,Martinez-Leon17}, here we exploit CS in the native optics hardware without modification to the underlying interferometer structure nor addition of new hardware components. Second, this is the first work to realize compression in the \emph{generalized interferometry} framework introduced recently by some of the authors to carry out modal analysis in arbitrary domains \cite{Abouraddy12OL,Lane17SR}. Third, we achieve compression gains both in the number of measurements used for reconstruction and the acquisition time. This is in contrast to prior work \cite{baha_OCT} employing CS in optical interferometry, which solely focused on reducing the number of measurements. Fourth, we establish analytical guarantees for successful modal reconstruction under sensing constraints set by the limited degrees of freedom of an optical interferometer.

We introduce CS into a prototypical optical system -- a two-path interferometer -- and demonstrate experimentally that sampling the interferogram below the Nyquist rate still enables spectral/modal recovery. We first cast the procedure of interferometric spectral/modal reconstruction as a linear measurement problem. We find that the reduced sensing matrix associated with random sub-Nyquist sampling satisfies sub-optimal conditions of isotropy and incoherence, thereby enabling compressive reconstruction from fewer samples \cite{Mardani15OE}. We denote this overall optical measurement scheme \textit{compressive interferometry} since it combines reduced data acquisition (reduced number of interferometric points sampled) with compressive coding (replacing harmonic analysis with CS reconstruction).

\begin{figure*}[t!]
\centering
\includegraphics[scale=1.1]{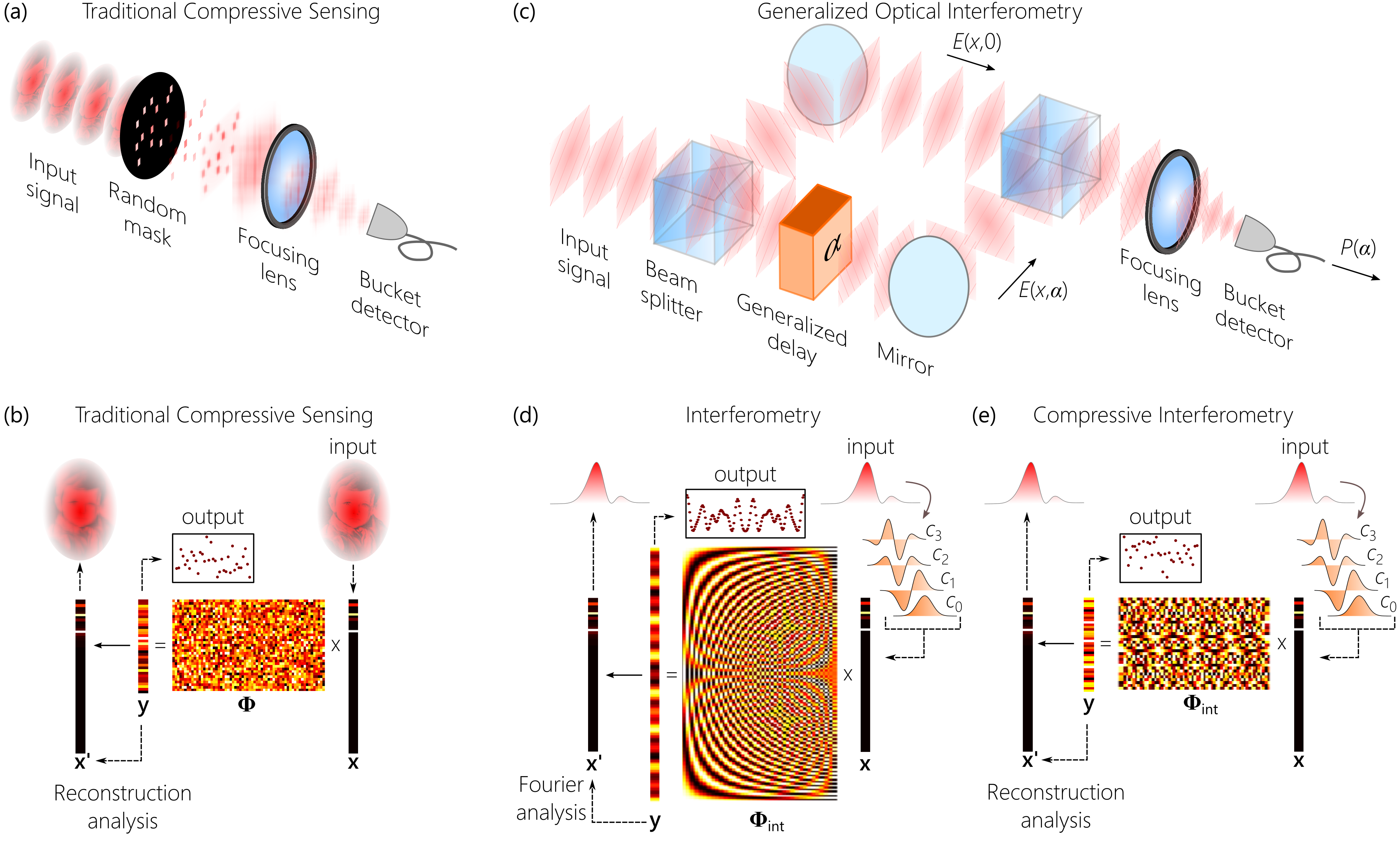}
\caption{\small \textbf{Concept of compressive optical interferometry. } \\
(a) Schematic of the traditional CS scheme in optics, where the field $E(x)$ is subject to random projections before measurements. (b) The mathematical concept used in traditional CS. The field is represented by a $M\times1$ vector $\mathbf{x}$ that is transformed by a $M\times N$ sensing matrix $\mathbf{\Phi}$ (with disordered entries corresponding to the random masks) to yield a $M\times1$ measurement vector $\mathbf{y}$, from which $\mathbf{x}$ is reconstructed. (c) The compressive interferometry scheme. The input beam $E(x)$ is directed to a Mach-Zhender interferferometer in which a generalized delay $\alpha$ replaces the usual temporal delay. Two copies of $E(x)$ are created at a beam splitter and the modes $\{\psi_{n}(x)\}$ underlying the beam acquire phase shifts of the form $e^{in\alpha}$ after passing through the generalized delay to yield a new beam $E(x;\alpha)$. The original and ‘delayed’ beams combine at another beam splitter, and their superposition is integrated by a bucket detector to produce an interferogram. (d,e) Graphical depiction of the matrix form of a generalized interferogram. (d) With evenly spaced Nyquist-rate sampling of the interferogram $M\geq2N$, the interferometric sensing matrix $\mathbf{\Phi}_{\mathrm{int}}$ has a well-defined deterministic structure. (e) With sub-Nyquist randomly sampled points $\alpha$, $\mathbf{\Phi}_{\mathrm{int}}$ now appears disordered and similar to $\mathbf{\Phi}$ in (b).}
\label{fig:concept} 
\end{figure*}

For sake of generality, we employ a two-path interferometer in which the usual temporal delay is replaced by a `generalized delay' that operates in the Hilbert space defined by any modal basis of interest \cite{Abouraddy11OL,Abouraddy12OL}. Such a delay corresponds to an optical transformation that has this modal basis as eigenfunctions, which reduces to the traditional temporal delay when spectral harmonics are of interest. The `delays' for Hermite-Gaussian (HG) or radial Laguerre-Gaussian (LG) modal bases are optical realizations of the fractional Fourier transform (FrFT) \cite{NAMIAS80JAM1,Ozaktas01wiley} or fractional Hankel transform (FrHT) \cite{NAMIAS80JAM2}, respectively. In all cases, the compressive interferometry strategy is to replace harmonic analysis of the Nyquist-sampled interferogram with random sub-Nyquist sampling \textit{without} modifying the native optical hardware. By constructing an intrinsically stable, basis-neutral interferometer that includes realizations of fractional transforms \cite{Lane17SR}, we confirm that CS helps successfully reconstruct the modal weights of sparse optical beams in a basis of HG or LG modes from the random sub-Nyquist samples. This may help in delay-intolerant applications that require real-time acquisition and processing, such as real-time imaging and high-speed communications exploiting spatial multiplexing.

 A typical optical implementation of CS is depicted in Fig.~\ref{fig:concept}(a)  \cite{Duarte08SPM, Howland14PRL, Mirhosseini14PRL}. Consider an optical field $E(x)=\sum_{n}c_{n}\psi_{n}(x)$ represented in an $N$-dimensional orthonormal basis $\{\psi_{n}(x)\}$, and $E(x)$ is normalized such $\sum|c_{n}|^{2}=1$. Here $x$ may refer to one or more spatial dimensions, or even time. The $N\times1$-vector $\mathbf{x}=\{c_{n}\}$ of coefficients is $s$-sparse if it receives contributions from $\leq s$ elements. In lieu of the $M\geq2N$ Nyquist-rate measurements normally required to reconstruct $\mathbf{x}$, CS leverages sparsity to reduce $M$ by projecting the field onto a sequence of \textit{random masks} selected from a Gaussian ensemble to produce a new $M\times1$ \textit{measurement vector} $\mathbf{y}$. Effectively, the field has undergone a linear transformation represented by a $M\times N$ \textit{sensing matrix} $\mathbf{\Phi}$, $\mathbf{y}=\mathbf{\Phi}\mathbf{x}$ [Fig.~\ref{fig:concept}(b)]. CS exploits this under-determined linear system to compressively recover $\mathbf{x}$ from $M\approx\mathcal{O}(s\log{N})\ll N$ linear measurements of $\mathbf{y}$ -- provided some conditions on $\mathbf{\Phi}$ are satisfied. For example, Basis Pursuit (BP) recovers $\mathbf{x}$ if $\mathbf{\Phi}$ satisfies the Restricted Isometry Property (RIP) \cite{Candes08CR}, which requires that $(1-\delta)\|\mathbf{x}\|^2\leq\|\mathbf{\Phi}\mathbf{x}\|^2\leq (1+\delta)\|\mathbf{x}\|^2$, for any $s$-sparse $\mathbf{x}$, where $0<\delta<1$ is a constant. Sensing matrices from Gaussian ensembles, for example, satisfy the RIP with overwhelming probability \cite{Candes08CR}. The random masks introduce new components into the optical system, thus adding to its complexity while reducing the number of measurements acquired.

The \textit{compressive interferometry} scheme relies on a different strategy. Consider an interferogram $P(\alpha)$ traced by scanning a `delay' $\alpha$ placed in one arm of a two-path interferometer [Fig.~\ref{fig:concept}(c)]. With respect to the modal basis $\{\psi_{n}(x)\}$, such a delay is an optical transformation having a continuous real control parameter $\alpha$, $\Lambda(x,x';\alpha)=\sum_{n}e^{in\alpha}\psi_{n}(x)\psi_{n}^{*}(x')$, which is a fractional transform having the basis functions as eigenstates with eigenvalues $e^{in\alpha}$, and $\Lambda(x,x';0)=\delta(x-x')$ \cite{Abouraddy12OL}. The parameter $\alpha$ acts as a `delay' in the Hilbert space spanned by $\{\psi_{n}(x)\}$ just as a delay $\tau$ does in the time domain. That $\alpha$ is in fact a delay can be readily seen by superposing the delayed field $E(x;\alpha)=\int dx\Lambda(x,x';\alpha)E(x')=\sum_{n}c_{n}e^{in\alpha}\psi_{n}(x)$ and the reference $E(x;0)$ to produce the interferogram
\begin{equation}\label{eq:interf_time}
P(\alpha)\propto\int dx|E(x;\alpha)+E(x;0)|^{2}=1+\sum_{n=1}^{N}|c_{n}|^{2}\cos(n\alpha).
\end{equation}
The modal coefficients $\mathbf{x}=\{|c_{n}|^{2}\}$ are revealed by taking a Fourier transform (FT) with respect to $\alpha$. This requires sampling $\alpha$ at the Nyquist rate dictated by the highest-order mode $N$ -- even if only a few modes contribute significantly [Fig.\ref{fig:concept}~(d)].

To show that CS algorithms may recover the sparse vector $\mathbf{x}$ from $M\ll N$ measurements, we first cast $P(\alpha)$ from Eq.~\ref{eq:interf_time} with \textit{selected values} of $\alpha$ into a linear model. A $M\times N$ interferometric sensing matrix $\mathbf{\Phi}_{\mathrm{int}}$ relates the $N\times1$ vector $\mathbf{x}$ of contributions from each Hilbert-space basis-element to the $M\times1$ measurement-vector $\mathbf{y}$ that contains the sampled interferogram [Fig.~\ref{fig:concept}(e)],
\begin{equation}\label{MatrixEquation}
\underbrace{\left[\begin{array}{c}P(\alpha_{1})-1\\P(\alpha_{2})-1\\\vdots\\P(\alpha_{M})-1\end{array}\right]}_{\mathbf{y}}\!\!\!=\!\!\!
\underbrace{\left[\begin{array}{cccc}\cos{\alpha_{1}}&\cos{2\alpha_{1}}&\ldots&\cos{N\alpha_{1}}\\
\cos{\alpha_{2}}&\cos{2\alpha_{2}}&\ldots&\cos{N\alpha_{2}}\\
	\vdots&\vdots&\ddots&\vdots\\
\cos{\alpha_{M}}&\cos{2\alpha_{M}}&\ldots&\cos{N\alpha_{M}}\end{array}\right]}_{\mathbf{\Phi}_{\mathrm{int}}}\!\!\!
\underbrace{\left[\begin{array}{c}|c_{1}|^{2}\\|c_{2}|^{2}\\\vdots\\|c_{N}|^{2}\end{array}\right]}_{\mathbf{x}}.
\end{equation}
Here $\mathbf{\Phi}_{\mathrm{int}}$ corresponds to \textit{any} interferometric measurement, temporal or otherwise. This matrix differs fundamentally from those typically employed in CS applications in optics [Fig.~\ref{fig:concept}(a) and \ref{fig:concept}(b)] where controllably random transformations $\mathbf{\Phi}$ are realized by an array of random patterns that are judiciously selected such that $\mathbf{\Phi}$ has the RIP, for example. In interferometry, this freedom in designing $\mathbf{\Phi}_{\mathrm{int}}$ is not available because $\mathbf{\Phi}_{\mathrm{int}}$ has a \textit{fixed structure} with few controllable degrees of freedom (the values of the delay parameter $\alpha$) imposed by the constrained sensing structure (the interferometer itself).

\begin{figure}[t!]
\centering
\includegraphics[scale=1]{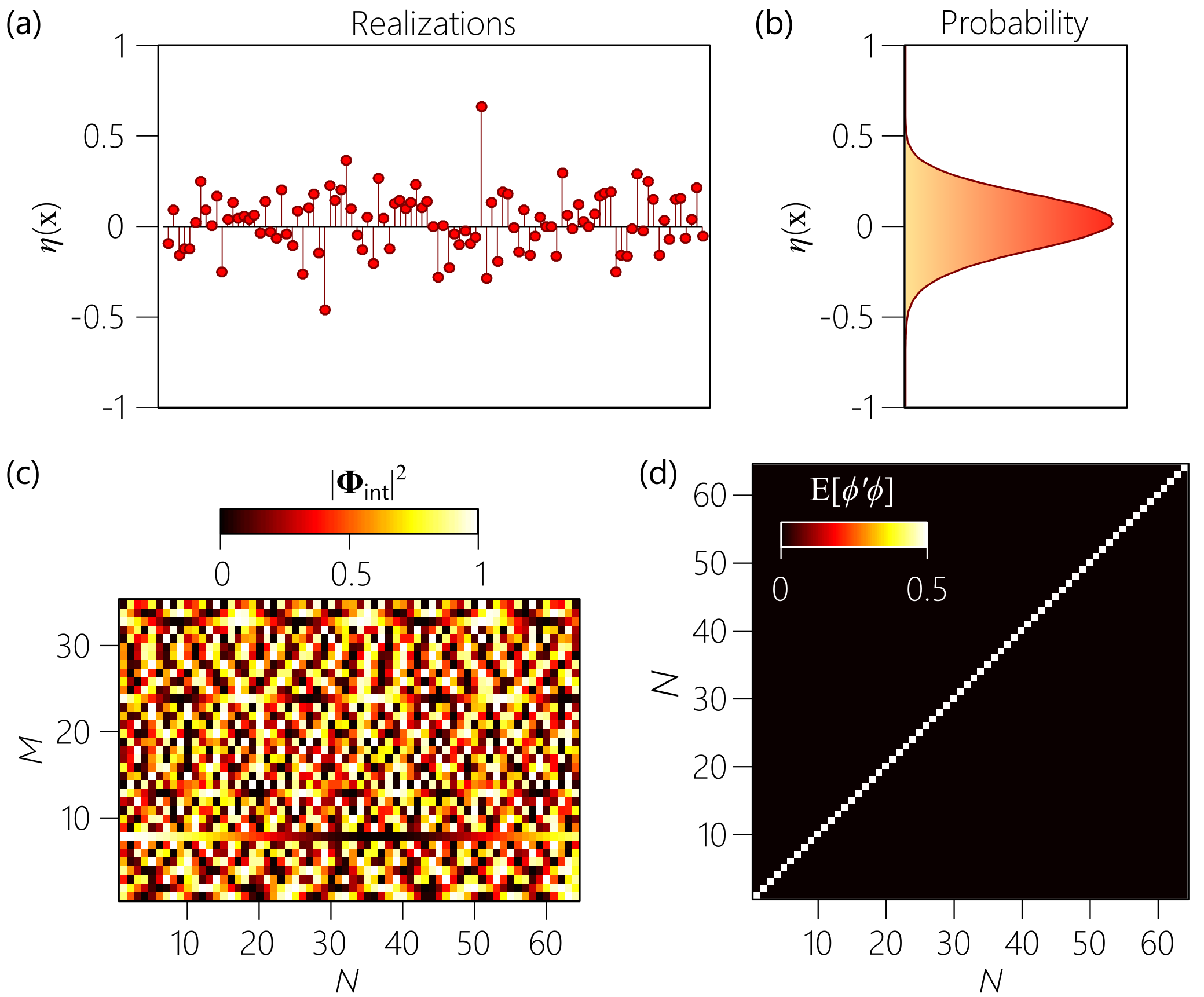}
\caption{\small \textbf{Restricted Isometry Property.}
\\(a) Calculated values of $\eta(\mathbf{x})=\|(\sqrt{2/M})\mathbf{\Phi}_{\mathrm{int}}\mathbf{x}\|^2/\|\mathbf{x}\|^2-1$ for $s$-sparse ($s=4$) vectors $\mathbf{x}$. (b) Histogram of $\eta(\mathbf{x})$ for an ensemble of $10^{6}$ realizations of $s$-sparse vectors. (c) The sensing matrix $\mathbf{\Phi}_{\mathrm{int}}$ from Eq.~\ref{MatrixEquation} with randomly sampled $\alpha$. (d) $\mathbf{\Phi}_{\mathrm{int}}$ is \textit{incoherent} with incoherence parameter $\mu=1$ and satisfies the \textit{isotropy} property $\mathrm{E}\{\boldsymbol{\phi}^{\dag}\boldsymbol{\phi}\}=0.5\,\mathbf{I}$.}
\label{fig:sensmatrix_properties}
\end{figure}

Given the constrained form of $\mathbf{\Phi}_{\mathrm{int}}$ in Eq.~\ref{MatrixEquation}, it can be shown that selecting random values of $\alpha$ from a uniform distribution at a sub-Nyquist rate yields $\mathbf{\Phi}_{\mathrm{int}}$ that satisfies the RIP with high probability when $M$ and $N$ are sufficiently large. For smaller values of $M$ and $N$, however, the RIP constant $\delta$ is large. 
The sufficient conditions for all existing reconstruction algorithms require smaller values of $\delta$ to guarantee successful reconstruction. 
For example, a sufficient condition for BP requires $\delta<\sqrt{2}-1$. 
Having the large $\delta$ associated with $\mathbf{\Phi}_{\mathrm{int}}$ therefore does not guarantee reconstruction on the basis of these requirements. 
As an illustration, in Fig.~\ref{fig:sensmatrix_properties}(a) we plot $\eta(\mathbf{x})=(\|(\sqrt{2/M})\mathbf{\Phi}_{\mathrm{int}}\mathbf{x}\|^2/\|\mathbf{x}\|^2)-1$ with a typical realization of the sensing matrix for a subset of an ensemble of $10^6$ sparse-vector realizations $\mathbf{x}$ with $s=4$, where $\max_\mathbf{x} |\eta(\mathbf{x})| \leq \delta$. A histogram constructed from the full ensemble is plotted in Fig.~\ref{fig:sensmatrix_properties}(b). However, the sensing matrix $\mathbf{\Phi}_{\mathrm{int}}$ with randomly selected $\alpha$ indeed belongs to a random ensemble satisfying weaker \textit{incoherence} and \textit{isotropy} properties despite its limited degrees of freedom imposed by the interferometric process. Provably, such matrices also yield exact reconstruction with high probability via BP \cite{candesRIPless}.

To clarify, let $\boldsymbol{\phi}$ denote a row of the sensing matrix selected from a random ensemble $\cal F$. The \textit{incoherence} parameter $\mu(\cal{F})$ is defined as the smallest number for which $\mathrm{max}_{n}|\phi_n|^2\leq\mu({\cal F})$, $n=1,2,\ldots,N$,
where $\phi_{n}$ are the entries of $\boldsymbol{\phi}$. The distribution $\cal{F}$ is said to obey the \textit{isotropy} property if 
$\mathrm{E}[\boldsymbol{\phi}^{\dag}\boldsymbol{\phi}]=a\mathbf{I}$, where $\mathrm{E}[.]$ is the expected value, $\mathbf{I}$ the identity matrix, $a$ some constant, and $\boldsymbol{\phi}^{\dag}$ the conjugate transpose of $\boldsymbol{\phi}$. $\mathbf{\Phi}_{\mathrm{int}}$ satisfies the conditions of incoherence and isotropy for $\alpha$ randomly selected from a uniform distribution over $[0,2\pi]$ with $\mu=1$, as confirmed in Fig.~\ref{fig:sensmatrix_properties}(c) that shows $\mathbf{\Phi}_{\mathrm{int}}$ consisting of $M$ rows selected from the full matrix in Fig.~\ref{fig:concept}(d). The matrix appears random in spite of being sampled from a structured sensing system. The matrix $\mathbf{\Phi}_{\mathrm{int}}$ is shown to satisfy the isotropy property in Fig.~\ref{fig:sensmatrix_properties}(d).

\begin{figure*}[t!]
\centering
\includegraphics[scale=1]{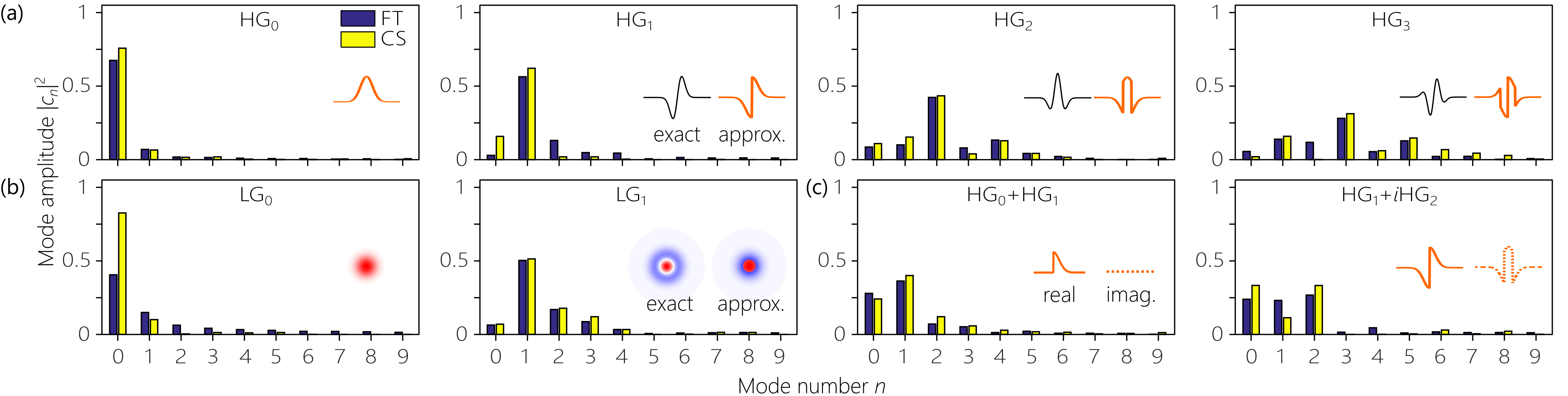}
\caption{\small\textbf{ Reconstructed modal distributions using the FT and CS approaches.} 
\\(a) The modal weights $|c_{n}|^2$ calculated for fields in the HG-basis by applying the FT to Nyquist-rate evenly sampled interferograms (blue) and CS to sub-Nyquist randomly sampled (yellow) interferograms. (b) Same as (a) for LG modes (LG$_0$ and LG$_1$) and (c) for superposition of HG modes (HG$_{0}+$HG$_1$ and HG$_{1}+i$HG$_2$). For FT, we use $128$ evenly spaced values of $\alpha$ from $0$ to $2\pi$. For CS, we use $M=30$ randomly selected $\alpha\sim{\cal U}[0,2\pi]$. The insets show the ideal (exact) and approximated modes that are implemented experimentally. For HG$_0$ and LG$_0$, the modes are produced without approximation.}
\label{fig:comp}
\end{figure*}

We now move on to implementing the compressive interferometric scheme experimentally. The generalized delay transformation $\Lambda(x,x';\alpha)$ has the modal basis $\{\psi_{n}(x)\}$ as eigenfunctions and its order parameter $\alpha$ represents the delay in the Hilbert space spanned by this basis. We carry out our experiments using two modal bases, 1D HG functions and radial LG functions. The delay transformations $\Lambda(x,x';\alpha)$ in these two modal bases correspond to the FrFT \cite{NAMIAS80JAM1,Ozaktas01wiley} and the FrHT \cite{NAMIAS80JAM2}, respectively \cite{Abouraddy11OL,Abouraddy12OL}. The usual temporal delay is replaced with an optical realization of the appropriate fractional transform, whose order $\alpha$ is swept in the interval $[0,2\pi]$ to produce an interferogram $P(\alpha)$. We construct these fractional transforms using spatial light modulators (SLMs) that realize generalized cylindrical or spherical lenses to produce the FrFT or FrHT, respectively, of desired order \cite{Rodrigo09OE}. We then exploit the polarization-selectivity of liquid-crystal-based SLMs \cite{Kagalwala13NP} to create an intrinsically stable common-path interferometer. One polarization -- that undergoes the transformation implemented by the SLMs -- is subjected to the generalized `delay' while the orthogonal polarization component is left invariant after the SLMs and thus corresponds to a reference \cite{Lane17SR}. A polarization projection at the output can then superpose the reference and the delayed beams, thereby yielding the target interferogram $P(\alpha)$ as the delay parameter $\alpha$ is swept evenly at the Nyquist rate.

\begin{figure}[t!]
\centering
\includegraphics[scale=1]{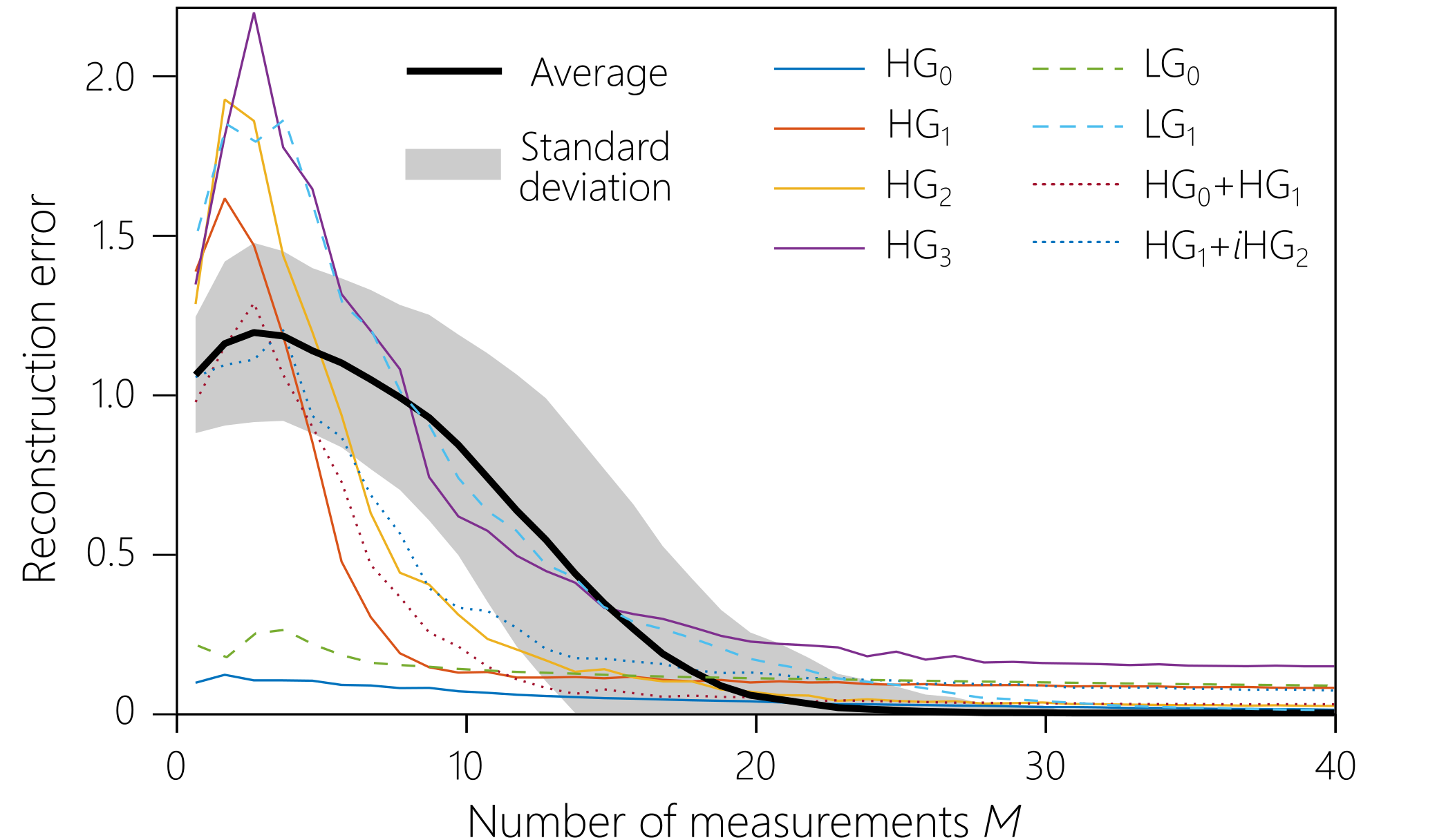}
\caption{\small \textbf{ Reconstruction error.}  \\ Reconstruction error  versus number of measurements $M$. Each curve results from averaging over $100$ runs of the experiment. The average curve (solid black) is the mean and the shaded area designate one standard deviation spread on either side of the required $M$ for $1000$ randomly generated examples of sparse vectors with support size $s\leq4$.}
\label{fig:error}
\end{figure}

In the compressive interferometry approach, $\alpha$ is randomly sampled at a sub-Nyquist rate. We study the performance of this strategy by comparing the modal reconstruction in the HG and radial LG bases to that obtained via the FT. $M$ measurements are collected randomly by selecting generalized phases $\alpha_{j}$, $j=1,2,\ldots,M$ from a uniform random distribution $[0,2\pi]$, and we use the BP algorithm to reconstruct the vector $\tilde{\mathbf{x}}$ of modal coefficients. The reconstructed vector is then compared to the vector $\mathbf{x}_{\mathrm{FT}}$ obtained from the FT of $128$ uniformly sampled interferometric measurements by computing the scaled error metric, $||\mathbf{x}_{\mathrm{FT}}-\tilde{\mathbf{x}}||^2/||\mathbf{x}_{\mathrm{FT}}||^2$. For each example, the average reconstruction error is evaluated by averaging over $100$ runs. The number of potential modes is $N=64$. Figures~\ref{fig:comp}(a) and \ref{fig:comp}(b) show the reconstructed modal coefficients using FT and CS for input beams formed from HG or LG modes. In Fig.~\ref{fig:comp}(c), we depict the reconstructed modal coefficients for two different superpositions of HG modes. The performance of the BP algorithm is comparable to that of FT using $M=30$ compressive measurements ($\approx25\%$ of the measurements in FT), leading to substantial savings in acquisition time without any added complexity to the system. The reconstruction error is plotted in Fig.~\ref{fig:error} as a function of $M$. The average error approaches zero when $M\approx25$. Statistical analysis of the observed error reveals that the required number of measurements lies between $M=15$ to $M=30$ to attain nearly error-free reconstruction.

In conclusion, we have devised a compressive approach to interferometry whereby the modal content of light beams can be recovered by sampling intereferograms randomly at sub-Nyquist rate using native optics hardware, which was demonstrated experimentally. Two fundamental differences distinguish our strategy from prior work employing CS in optics. First, the interferometric sensing matrix is imposed through the structure of the interferometer. As such, compression has to be carried out under sensing constraints enforced by the limited degrees of freedom of the sensing system. Second, previous work in the context of interferometry focused on reducing the number of measurements used for reconstruction, but not on \emph{compressive data acquisition}. Leveraging the notion of generalized delay, our approach enables compressive modal analysis in arbitrary modal basis. Furthermore, we established provable guarantees for successful reconstruction despite the sensing constraints set by the limited degrees of freedom of the interferometer.

\vspace{3mm}
\noindent \textbf{Funding.}  Office of Naval Research (ONR) contracts N00014-14-1-0260 and N00014-17-1-2458.

\newpage 
\bibliography{references}

\begin{thebibliography}{31}%
\makeatletter
\providecommand \@ifxundefined [1]{%
 \@ifx{#1\undefined}
}%
\providecommand \@ifnum [1]{%
 \ifnum #1\expandafter \@firstoftwo
 \else \expandafter \@secondoftwo
 \fi
}%
\providecommand \@ifx [1]{%
 \ifx #1\expandafter \@firstoftwo
 \else \expandafter \@secondoftwo
 \fi
}%
\providecommand \natexlab [1]{#1}%
\providecommand \enquote  [1]{``#1''}%
\providecommand \bibnamefont  [1]{#1}%
\providecommand \bibfnamefont [1]{#1}%
\providecommand \citenamefont [1]{#1}%
\providecommand \href@noop [0]{\@secondoftwo}%
\providecommand \href [0]{\begingroup \@sanitize@url \@href}%
\providecommand \@href[1]{\@@startlink{#1}\@@href}%
\providecommand \@@href[1]{\endgroup#1\@@endlink}%
\providecommand \@sanitize@url [0]{\catcode `\\12\catcode `\$12\catcode
  `\&12\catcode `\#12\catcode `\^12\catcode `\_12\catcode `\%12\relax}%
\providecommand \@@startlink[1]{}%
\providecommand \@@endlink[0]{}%
\providecommand \url  [0]{\begingroup\@sanitize@url \@url }%
\providecommand \@url [1]{\endgroup\@href {#1}{\urlprefix }}%
\providecommand \urlprefix  [0]{URL }%
\providecommand \Eprint [0]{\href }%
\providecommand \doibase [0]{http://dx.doi.org/}%
\providecommand \selectlanguage [0]{\@gobble}%
\providecommand \bibinfo  [0]{\@secondoftwo}%
\providecommand \bibfield  [0]{\@secondoftwo}%
\providecommand \translation [1]{[#1]}%
\providecommand \BibitemOpen [0]{}%
\providecommand \bibitemStop [0]{}%
\providecommand \bibitemNoStop [0]{.\EOS\space}%
\providecommand \EOS [0]{\spacefactor3000\relax}%
\providecommand \BibitemShut  [1]{\csname bibitem#1\endcsname}%
\let\auto@bib@innerbib\@empty
\bibitem [{\citenamefont {Donoho}(2006)}]{Donoho06TIT}%
  \BibitemOpen
  \bibfield  {author} {\bibinfo {author} {\bibfnamefont {D.~L.}\ \bibnamefont
  {Donoho}},\ }\href@noop {} {\bibfield  {journal} {\bibinfo  {journal} {IEEE
  Trans. Inf. Theory}\ }\textbf {\bibinfo {volume} {52}},\ \bibinfo {pages}
  {1289} (\bibinfo {year} {2006})}\BibitemShut {NoStop}%
\bibitem [{\citenamefont {Candes}(2008)}]{Candes08CR}%
  \BibitemOpen
  \bibfield  {author} {\bibinfo {author} {\bibfnamefont {E.~J.}\ \bibnamefont
  {Candes}},\ }\href@noop {} {\bibfield  {journal} {\bibinfo  {journal} {C. R.
  Acad. Sci.}\ }\textbf {\bibinfo {volume} {346}},\ \bibinfo {pages} {589}
  (\bibinfo {year} {2008})}\BibitemShut {NoStop}%
\bibitem [{\citenamefont {Candes}(2006)}]{Candes06PICM}%
  \BibitemOpen
  \bibfield  {author} {\bibinfo {author} {\bibfnamefont {E.~J.}\ \bibnamefont
  {Candes}},\ }\href@noop {} {\bibfield  {journal} {\bibinfo  {journal} {Proc.
  Int. Congress of Mathematicians}\ } (\bibinfo {year} {2006})}\BibitemShut
  {NoStop}%
\bibitem [{\citenamefont {Duarte}\ \emph {et~al.}(2008)\citenamefont {Duarte},
  \citenamefont {Davenport}, \citenamefont {Takhar}, \citenamefont {Laska},
  \citenamefont {Sun}, \citenamefont {Kelly},\ and\ \citenamefont
  {Baraniuk}}]{Duarte08SPM}%
  \BibitemOpen
  \bibfield  {author} {\bibinfo {author} {\bibfnamefont {M.~F.}\ \bibnamefont
  {Duarte}}, \bibinfo {author} {\bibfnamefont {M.~A.}\ \bibnamefont
  {Davenport}}, \bibinfo {author} {\bibfnamefont {D.}~\bibnamefont {Takhar}},
  \bibinfo {author} {\bibfnamefont {J.~N.}\ \bibnamefont {Laska}}, \bibinfo
  {author} {\bibfnamefont {T.}~\bibnamefont {Sun}}, \bibinfo {author}
  {\bibfnamefont {K.~F.}\ \bibnamefont {Kelly}}, \ and\ \bibinfo {author}
  {\bibfnamefont {R.~G.}\ \bibnamefont {Baraniuk}},\ }\href@noop {} {\bibfield
  {journal} {\bibinfo  {journal} {IEEE Signal Process. Mag.}\ }\textbf
  {\bibinfo {volume} {25}},\ \bibinfo {pages} {83} (\bibinfo {year}
  {2008})}\BibitemShut {NoStop}%
\bibitem [{\citenamefont {Magalhaes}\ \emph {et~al.}(2011)\citenamefont
  {Magalhaes}, \citenamefont {Araujo}, \citenamefont {Correia}, \citenamefont
  {Abolbashari},\ and\ \citenamefont {Farahi}}]{Magalhaes11AO}%
  \BibitemOpen
  \bibfield  {author} {\bibinfo {author} {\bibfnamefont {F.}~\bibnamefont
  {Magalhaes}}, \bibinfo {author} {\bibfnamefont {F.~M.}\ \bibnamefont
  {Araujo}}, \bibinfo {author} {\bibfnamefont {M.~V.}\ \bibnamefont {Correia}},
  \bibinfo {author} {\bibfnamefont {M.}~\bibnamefont {Abolbashari}}, \ and\
  \bibinfo {author} {\bibfnamefont {F.}~\bibnamefont {Farahi}},\ }\href@noop {}
  {\bibfield  {journal} {\bibinfo  {journal} {Appl. Opt.}\ }\textbf {\bibinfo
  {volume} {50}},\ \bibinfo {pages} {405} (\bibinfo {year} {2011})}\BibitemShut
  {NoStop}%
\bibitem [{\citenamefont {Martinez-Leon}\ \emph {et~al.}(2017)\citenamefont
  {Martinez-Leon}, \citenamefont {Clemente}, \citenamefont {Mori},
  \citenamefont {Climent}, \citenamefont {Lancis},\ and\ \citenamefont
  {Tajahuerce}}]{Martinez-Leon17}%
  \BibitemOpen
  \bibfield  {author} {\bibinfo {author} {\bibfnamefont {L.}~\bibnamefont
  {Martinez-Leon}}, \bibinfo {author} {\bibfnamefont {P.}~\bibnamefont
  {Clemente}}, \bibinfo {author} {\bibfnamefont {Y.}~\bibnamefont {Mori}},
  \bibinfo {author} {\bibfnamefont {V.}~\bibnamefont {Climent}}, \bibinfo
  {author} {\bibfnamefont {J.}~\bibnamefont {Lancis}}, \ and\ \bibinfo {author}
  {\bibfnamefont {E.}~\bibnamefont {Tajahuerce}},\ }\href@noop {} {\bibfield
  {journal} {\bibinfo  {journal} {Opt. Express}\ }\textbf {\bibinfo {volume}
  {25}},\ \bibinfo {pages} {4975} (\bibinfo {year} {2017})}\BibitemShut
  {NoStop}%
\bibitem [{\citenamefont {Clemente}\ \emph {et~al.}(2013)\citenamefont
  {Clemente}, \citenamefont {Dur\'{a}n}, \citenamefont {Tajahuerce},
  \citenamefont {Andr\'{e}s}, \citenamefont {Climent},\ and\ \citenamefont
  {Lancis}}]{Clemente13}%
  \BibitemOpen
  \bibfield  {author} {\bibinfo {author} {\bibfnamefont {P.}~\bibnamefont
  {Clemente}}, \bibinfo {author} {\bibfnamefont {V.}~\bibnamefont {Dur\'{a}n}},
  \bibinfo {author} {\bibfnamefont {E.}~\bibnamefont {Tajahuerce}}, \bibinfo
  {author} {\bibfnamefont {P.}~\bibnamefont {Andr\'{e}s}}, \bibinfo {author}
  {\bibfnamefont {V.}~\bibnamefont {Climent}}, \ and\ \bibinfo {author}
  {\bibfnamefont {J.}~\bibnamefont {Lancis}},\ }\href@noop {} {\bibfield
  {journal} {\bibinfo  {journal} {Opt. Lett.}\ }\textbf {\bibinfo {volume}
  {38}},\ \bibinfo {pages} {2524} (\bibinfo {year} {2013})}\BibitemShut
  {NoStop}%
\bibitem [{\citenamefont {Deepan}\ \emph {et~al.}(2014)\citenamefont {Deepan},
  \citenamefont {Quan},\ and\ \citenamefont {Tay}}]{Deepan14SPIE}%
  \BibitemOpen
  \bibfield  {author} {\bibinfo {author} {\bibfnamefont {B.}~\bibnamefont
  {Deepan}}, \bibinfo {author} {\bibfnamefont {C.}~\bibnamefont {Quan}}, \ and\
  \bibinfo {author} {\bibfnamefont {C.~J.}\ \bibnamefont {Tay}},\ }\href@noop
  {} {\bibfield  {journal} {\bibinfo  {journal} {Proc. SPIE}\ }\textbf
  {\bibinfo {volume} {9234}},\ \bibinfo {pages} {923419} (\bibinfo {year}
  {2014})}\BibitemShut {NoStop}%
\bibitem [{\citenamefont {Rivenson}\ \emph {et~al.}(2013)\citenamefont
  {Rivenson}, \citenamefont {Stern},\ and\ \citenamefont {Rosen}}]{Yair13OL}%
  \BibitemOpen
  \bibfield  {author} {\bibinfo {author} {\bibfnamefont {Y.}~\bibnamefont
  {Rivenson}}, \bibinfo {author} {\bibfnamefont {A.}~\bibnamefont {Stern}}, \
  and\ \bibinfo {author} {\bibfnamefont {J.}~\bibnamefont {Rosen}},\
  }\href@noop {} {\bibfield  {journal} {\bibinfo  {journal} {Opt. Lett.}\
  }\textbf {\bibinfo {volume} {38}},\ \bibinfo {pages} {2509} (\bibinfo {year}
  {2013})}\BibitemShut {NoStop}%
\bibitem [{\citenamefont {Liutkus}\ \emph {et~al.}(2014)\citenamefont
  {Liutkus}, \citenamefont {Martina}, \citenamefont {Popoff}, \citenamefont
  {Chardon}, \citenamefont {Katz}, \citenamefont {Lerosey}, \citenamefont
  {Gigan}, \citenamefont {Daudet},\ and\ \citenamefont {Carron}}]{Liutkus14SR}%
  \BibitemOpen
  \bibfield  {author} {\bibinfo {author} {\bibfnamefont {A.}~\bibnamefont
  {Liutkus}}, \bibinfo {author} {\bibfnamefont {D.}~\bibnamefont {Martina}},
  \bibinfo {author} {\bibfnamefont {S.}~\bibnamefont {Popoff}}, \bibinfo
  {author} {\bibfnamefont {G.}~\bibnamefont {Chardon}}, \bibinfo {author}
  {\bibfnamefont {O.}~\bibnamefont {Katz}}, \bibinfo {author} {\bibfnamefont
  {G.}~\bibnamefont {Lerosey}}, \bibinfo {author} {\bibfnamefont
  {S.}~\bibnamefont {Gigan}}, \bibinfo {author} {\bibfnamefont
  {L.}~\bibnamefont {Daudet}}, \ and\ \bibinfo {author} {\bibfnamefont
  {I.}~\bibnamefont {Carron}},\ }\href@noop {} {\bibfield  {journal} {\bibinfo
  {journal} {Sci. Rep.}\ }\textbf {\bibinfo {volume} {4}},\ \bibinfo {pages}
  {5552} (\bibinfo {year} {2014})}\BibitemShut {NoStop}%
\bibitem [{\citenamefont {Graff}\ and\ \citenamefont
  {Sidky}(2015)}]{Graff15AO}%
  \BibitemOpen
  \bibfield  {author} {\bibinfo {author} {\bibfnamefont {C.~G.}\ \bibnamefont
  {Graff}}\ and\ \bibinfo {author} {\bibfnamefont {E.~Y.}\ \bibnamefont
  {Sidky}},\ }\href@noop {} {\bibfield  {journal} {\bibinfo  {journal} {Appl.
  Opt.}\ }\textbf {\bibinfo {volume} {54}},\ \bibinfo {pages} {C23} (\bibinfo
  {year} {2015})}\BibitemShut {NoStop}%
\bibitem [{\citenamefont {S\"{u}zen}\ \emph {et~al.}(2010)\citenamefont
  {S\"{u}zen}, \citenamefont {Giannoula},\ and\ \citenamefont
  {Durduran}}]{Mehmet10OE}%
  \BibitemOpen
  \bibfield  {author} {\bibinfo {author} {\bibfnamefont {M.}~\bibnamefont
  {S\"{u}zen}}, \bibinfo {author} {\bibfnamefont {A.}~\bibnamefont
  {Giannoula}}, \ and\ \bibinfo {author} {\bibfnamefont {T.}~\bibnamefont
  {Durduran}},\ }\href@noop {} {\bibfield  {journal} {\bibinfo  {journal} {Opt.
  Express}\ }\textbf {\bibinfo {volume} {18}},\ \bibinfo {pages} {23676}
  (\bibinfo {year} {2010})}\BibitemShut {NoStop}%
\bibitem [{\citenamefont {Yao}\ \emph {et~al.}(2015)\citenamefont {Yao},
  \citenamefont {Pian},\ and\ \citenamefont {Intes}}]{Yao15BOE}%
  \BibitemOpen
  \bibfield  {author} {\bibinfo {author} {\bibfnamefont {R.}~\bibnamefont
  {Yao}}, \bibinfo {author} {\bibfnamefont {Q.}~\bibnamefont {Pian}}, \ and\
  \bibinfo {author} {\bibfnamefont {X.}~\bibnamefont {Intes}},\ }\href@noop {}
  {\bibfield  {journal} {\bibinfo  {journal} {Biomed. Opt. Express}\ }\textbf
  {\bibinfo {volume} {6}},\ \bibinfo {pages} {4887} (\bibinfo {year}
  {2015})}\BibitemShut {NoStop}%
\bibitem [{\citenamefont {Li}\ \emph {et~al.}(2015)\citenamefont {Li},
  \citenamefont {Li}, \citenamefont {Pan},\ and\ \citenamefont
  {Li}}]{Li15SCIrep}%
  \BibitemOpen
  \bibfield  {author} {\bibinfo {author} {\bibfnamefont {J.}~\bibnamefont
  {Li}}, \bibinfo {author} {\bibfnamefont {J.~S.}\ \bibnamefont {Li}}, \bibinfo
  {author} {\bibfnamefont {Y.~Y.}\ \bibnamefont {Pan}}, \ and\ \bibinfo
  {author} {\bibfnamefont {R.}~\bibnamefont {Li}},\ }\href@noop {} {\bibfield
  {journal} {\bibinfo  {journal} {Sci. Rep.}\ }\textbf {\bibinfo {volume}
  {5}},\ \bibinfo {pages} {10374} (\bibinfo {year} {2015})}\BibitemShut
  {NoStop}%
\bibitem [{\citenamefont {Rawat}\ \emph {et~al.}(2015)\citenamefont {Rawat},
  \citenamefont {Kim}, \citenamefont {Muniraj}, \citenamefont {Situ},\ and\
  \citenamefont {Lee}}]{Rawat15AP}%
  \BibitemOpen
  \bibfield  {author} {\bibinfo {author} {\bibfnamefont {N.}~\bibnamefont
  {Rawat}}, \bibinfo {author} {\bibfnamefont {B.}~\bibnamefont {Kim}}, \bibinfo
  {author} {\bibfnamefont {I.}~\bibnamefont {Muniraj}}, \bibinfo {author}
  {\bibfnamefont {G.}~\bibnamefont {Situ}}, \ and\ \bibinfo {author}
  {\bibfnamefont {B.-G.}\ \bibnamefont {Lee}},\ }\href@noop {} {\bibfield
  {journal} {\bibinfo  {journal} {Appl. Opt.}\ }\textbf {\bibinfo {volume}
  {54}},\ \bibinfo {pages} {1782} (\bibinfo {year} {2015})}\BibitemShut
  {NoStop}%
\bibitem [{\citenamefont {Wang}\ and\ \citenamefont {Yu}(2014)}]{Zhu14OE}%
  \BibitemOpen
  \bibfield  {author} {\bibinfo {author} {\bibfnamefont {Z.}~\bibnamefont
  {Wang}}\ and\ \bibinfo {author} {\bibfnamefont {Z.}~\bibnamefont {Yu}},\
  }\href@noop {} {\bibfield  {journal} {\bibinfo  {journal} {Opt. Express}\
  }\textbf {\bibinfo {volume} {22}},\ \bibinfo {pages} {25608} (\bibinfo {year}
  {2014})}\BibitemShut {NoStop}%
\bibitem [{\citenamefont {Howland}\ \emph {et~al.}(2014)\citenamefont
  {Howland}, \citenamefont {Schneeloch}, \citenamefont {Lum},\ and\
  \citenamefont {Howell}}]{Howland14PRL}%
  \BibitemOpen
  \bibfield  {author} {\bibinfo {author} {\bibfnamefont {G.~A.}\ \bibnamefont
  {Howland}}, \bibinfo {author} {\bibfnamefont {J.}~\bibnamefont {Schneeloch}},
  \bibinfo {author} {\bibfnamefont {D.~J.}\ \bibnamefont {Lum}}, \ and\
  \bibinfo {author} {\bibfnamefont {J.~C.}\ \bibnamefont {Howell}},\
  }\href@noop {} {\bibfield  {journal} {\bibinfo  {journal} {Phys. Rev. Lett.}\
  }\textbf {\bibinfo {volume} {112}},\ \bibinfo {pages} {253602} (\bibinfo
  {year} {2014})}\BibitemShut {NoStop}%
\bibitem [{\citenamefont {Mirhosseini}\ \emph {et~al.}(2014)\citenamefont
  {Mirhosseini}, \citenamefont {Maga{\~n}a-Loaiza}, \citenamefont
  {Rafsanjani},\ and\ \citenamefont {Boyd}}]{Mirhosseini14PRL}%
  \BibitemOpen
  \bibfield  {author} {\bibinfo {author} {\bibfnamefont {M.}~\bibnamefont
  {Mirhosseini}}, \bibinfo {author} {\bibfnamefont {O.~S.}\ \bibnamefont
  {Maga{\~n}a-Loaiza}}, \bibinfo {author} {\bibfnamefont {S.~M.~H.}\
  \bibnamefont {Rafsanjani}}, \ and\ \bibinfo {author} {\bibfnamefont {R.~W.}\
  \bibnamefont {Boyd}},\ }\href@noop {} {\bibfield  {journal} {\bibinfo
  {journal} {Phys. Rev. Lett.}\ }\textbf {\bibinfo {volume} {113}},\ \bibinfo
  {pages} {090402} (\bibinfo {year} {2014})}\BibitemShut {NoStop}%
\bibitem [{\citenamefont {Kalev}\ \emph {et~al.}(2015)\citenamefont {Kalev},
  \citenamefont {Kosut},\ and\ \citenamefont {Deutsch}}]{Kalev15QI}%
  \BibitemOpen
  \bibfield  {author} {\bibinfo {author} {\bibfnamefont {A.}~\bibnamefont
  {Kalev}}, \bibinfo {author} {\bibfnamefont {R.~L.}\ \bibnamefont {Kosut}}, \
  and\ \bibinfo {author} {\bibfnamefont {I.~H.}\ \bibnamefont {Deutsch}},\
  }\href@noop {} {\bibfield  {journal} {\bibinfo  {journal} {npj Quantum Inf.}\
  }\textbf {\bibinfo {volume} {1}},\ \bibinfo {pages} {15018} (\bibinfo {year}
  {2015})}\BibitemShut {NoStop}%
\bibitem [{\citenamefont {Howland}\ \emph {et~al.}(2016)\citenamefont
  {Howland}, \citenamefont {Knarr}, \citenamefont {Schneeloch}, \citenamefont
  {Lum},\ and\ \citenamefont {Howell}}]{Howland16PRX}%
  \BibitemOpen
  \bibfield  {author} {\bibinfo {author} {\bibfnamefont {G.~A.}\ \bibnamefont
  {Howland}}, \bibinfo {author} {\bibfnamefont {S.~H.}\ \bibnamefont {Knarr}},
  \bibinfo {author} {\bibfnamefont {J.}~\bibnamefont {Schneeloch}}, \bibinfo
  {author} {\bibfnamefont {D.~J.}\ \bibnamefont {Lum}}, \ and\ \bibinfo
  {author} {\bibfnamefont {J.~C.}\ \bibnamefont {Howell}},\ }\href@noop {}
  {\bibfield  {journal} {\bibinfo  {journal} {Phys. Rev. X}\ }\textbf {\bibinfo
  {volume} {6}},\ \bibinfo {pages} {021018} (\bibinfo {year}
  {2016})}\BibitemShut {NoStop}%
\bibitem [{\citenamefont {Abouraddy}\ \emph {et~al.}(2012)\citenamefont
  {Abouraddy}, \citenamefont {Yarnall},\ and\ \citenamefont
  {Saleh}}]{Abouraddy12OL}%
  \BibitemOpen
  \bibfield  {author} {\bibinfo {author} {\bibfnamefont {A.~F.}\ \bibnamefont
  {Abouraddy}}, \bibinfo {author} {\bibfnamefont {T.~M.}\ \bibnamefont
  {Yarnall}}, \ and\ \bibinfo {author} {\bibfnamefont {B.~E.~A.}\ \bibnamefont
  {Saleh}},\ }\href@noop {} {\bibfield  {journal} {\bibinfo  {journal} {Opt.
  Lett.}\ }\textbf {\bibinfo {volume} {37}},\ \bibinfo {pages} {2889} (\bibinfo
  {year} {2012})}\BibitemShut {NoStop}%
\bibitem [{\citenamefont {Martin}\ \emph {et~al.}(2017)\citenamefont {Martin},
  \citenamefont {Mardani}, \citenamefont {Kondakci}, \citenamefont {Larson},
  \citenamefont {Shabahang}, \citenamefont {Jahromi}, \citenamefont {Malhotra},
  \citenamefont {Vamivakas}, \citenamefont {Atia},\ and\ \citenamefont
  {Abouraddy}}]{Lane17SR}%
  \BibitemOpen
  \bibfield  {author} {\bibinfo {author} {\bibfnamefont {L.}~\bibnamefont
  {Martin}}, \bibinfo {author} {\bibfnamefont {D.}~\bibnamefont {Mardani}},
  \bibinfo {author} {\bibfnamefont {H.~E.}\ \bibnamefont {Kondakci}}, \bibinfo
  {author} {\bibfnamefont {W.~D.}\ \bibnamefont {Larson}}, \bibinfo {author}
  {\bibfnamefont {S.}~\bibnamefont {Shabahang}}, \bibinfo {author}
  {\bibfnamefont {A.~K.}\ \bibnamefont {Jahromi}}, \bibinfo {author}
  {\bibfnamefont {T.}~\bibnamefont {Malhotra}}, \bibinfo {author}
  {\bibfnamefont {A.~N.}\ \bibnamefont {Vamivakas}}, \bibinfo {author}
  {\bibfnamefont {G.~K.}\ \bibnamefont {Atia}}, \ and\ \bibinfo {author}
  {\bibfnamefont {A.~F.}\ \bibnamefont {Abouraddy}},\ }\href@noop {} {\bibfield
   {journal} {\bibinfo  {journal} {Sci. Rep.}\ }\textbf {\bibinfo {volume}
  {7}},\ \bibinfo {pages} {44995} (\bibinfo {year} {2017})}\BibitemShut
  {NoStop}%
\bibitem [{\citenamefont {Mohan}\ \emph {et~al.}(2010)\citenamefont {Mohan},
  \citenamefont {Stojanovic}, \citenamefont {Karl}, \citenamefont {Saleh},\
  and\ \citenamefont {Teich}}]{baha_OCT}%
  \BibitemOpen
  \bibfield  {author} {\bibinfo {author} {\bibfnamefont {N.}~\bibnamefont
  {Mohan}}, \bibinfo {author} {\bibfnamefont {I.}~\bibnamefont {Stojanovic}},
  \bibinfo {author} {\bibfnamefont {W.~C.}\ \bibnamefont {Karl}}, \bibinfo
  {author} {\bibfnamefont {B.~E.~A.}\ \bibnamefont {Saleh}}, \ and\ \bibinfo
  {author} {\bibfnamefont {M.~C.}\ \bibnamefont {Teich}},\ }\href@noop {}
  {\bibfield  {journal} {\bibinfo  {journal} {Proc. SPIE}\ }\textbf {\bibinfo
  {volume} {7570}},\ \bibinfo {pages} {75700L} (\bibinfo {year}
  {2010})}\BibitemShut {NoStop}%
\bibitem [{\citenamefont {Mardani}\ \emph {et~al.}(2015)\citenamefont
  {Mardani}, \citenamefont {Abouraddy},\ and\ \citenamefont
  {Atia}}]{Mardani15OE}%
  \BibitemOpen
  \bibfield  {author} {\bibinfo {author} {\bibfnamefont {D.}~\bibnamefont
  {Mardani}}, \bibinfo {author} {\bibfnamefont {A.~F.}\ \bibnamefont
  {Abouraddy}}, \ and\ \bibinfo {author} {\bibfnamefont {G.~K.}\ \bibnamefont
  {Atia}},\ }\href@noop {} {\bibfield  {journal} {\bibinfo  {journal} {Opt.
  Express}\ }\textbf {\bibinfo {volume} {23}},\ \bibinfo {pages} {28449}
  (\bibinfo {year} {2015})}\BibitemShut {NoStop}%
\bibitem [{\citenamefont {Abouraddy}\ \emph {et~al.}(2011)\citenamefont
  {Abouraddy}, \citenamefont {Yarnall},\ and\ \citenamefont
  {Saleh}}]{Abouraddy11OL}%
  \BibitemOpen
  \bibfield  {author} {\bibinfo {author} {\bibfnamefont {A.~F.}\ \bibnamefont
  {Abouraddy}}, \bibinfo {author} {\bibfnamefont {T.~M.}\ \bibnamefont
  {Yarnall}}, \ and\ \bibinfo {author} {\bibfnamefont {B.~E.~A.}\ \bibnamefont
  {Saleh}},\ }\href@noop {} {\bibfield  {journal} {\bibinfo  {journal} {Opt.
  Lett.}\ }\textbf {\bibinfo {volume} {36}},\ \bibinfo {pages} {4683} (\bibinfo
  {year} {2011})}\BibitemShut {NoStop}%
\bibitem [{\citenamefont {Namias}(1980{\natexlab{a}})}]{NAMIAS80JAM1}%
  \BibitemOpen
  \bibfield  {author} {\bibinfo {author} {\bibfnamefont {V.}~\bibnamefont
  {Namias}},\ }\href@noop {} {\bibfield  {journal} {\bibinfo  {journal} {IMA J.
  Appl. Math.}\ }\textbf {\bibinfo {volume} {25}},\ \bibinfo {pages} {241}
  (\bibinfo {year} {1980}{\natexlab{a}})}\BibitemShut {NoStop}%
\bibitem [{\citenamefont {Ozaktas}\ \emph {et~al.}(2001)\citenamefont
  {Ozaktas}, \citenamefont {Zalevsky},\ and\ \citenamefont
  {Kutay}}]{Ozaktas01wiley}%
  \BibitemOpen
  \bibfield  {author} {\bibinfo {author} {\bibfnamefont {H.~M.}\ \bibnamefont
  {Ozaktas}}, \bibinfo {author} {\bibfnamefont {Z.}~\bibnamefont {Zalevsky}}, \
  and\ \bibinfo {author} {\bibfnamefont {M.~A.}\ \bibnamefont {Kutay}},\
  }\href@noop {} {\emph {\bibinfo {title} {The Fractional {F}ourier Transform
  with Applications in Optics and Signal Processing}}}\ (\bibinfo  {publisher}
  {Wiley},\ \bibinfo {year} {2001})\BibitemShut {NoStop}%
\bibitem [{\citenamefont {Namias}(1980{\natexlab{b}})}]{NAMIAS80JAM2}%
  \BibitemOpen
  \bibfield  {author} {\bibinfo {author} {\bibfnamefont {V.}~\bibnamefont
  {Namias}},\ }\href@noop {} {\bibfield  {journal} {\bibinfo  {journal} {IMA J.
  Appl. Math.}\ }\textbf {\bibinfo {volume} {26}},\ \bibinfo {pages} {187}
  (\bibinfo {year} {1980}{\natexlab{b}})}\BibitemShut {NoStop}%
\bibitem [{\citenamefont {Candes}\ and\ \citenamefont
  {Plan}(2011)}]{candesRIPless}%
  \BibitemOpen
  \bibfield  {author} {\bibinfo {author} {\bibfnamefont {E.~J.}\ \bibnamefont
  {Candes}}\ and\ \bibinfo {author} {\bibfnamefont {Y.}~\bibnamefont {Plan}},\
  }\href@noop {} {\bibfield  {journal} {\bibinfo  {journal} {IEEE Trans. Inf.
  Theory}\ }\textbf {\bibinfo {volume} {57}},\ \bibinfo {pages} {7235}
  (\bibinfo {year} {2011})}\BibitemShut {NoStop}%
\bibitem [{\citenamefont {Rodrigo}\ \emph {et~al.}(2009)\citenamefont
  {Rodrigo}, \citenamefont {Alieva},\ and\ \citenamefont
  {Calvo}}]{Rodrigo09OE}%
  \BibitemOpen
  \bibfield  {author} {\bibinfo {author} {\bibfnamefont {J.~A.}\ \bibnamefont
  {Rodrigo}}, \bibinfo {author} {\bibfnamefont {T.}~\bibnamefont {Alieva}}, \
  and\ \bibinfo {author} {\bibfnamefont {M.~L.}\ \bibnamefont {Calvo}},\
  }\href@noop {} {\bibfield  {journal} {\bibinfo  {journal} {Opt. Express}\
  }\textbf {\bibinfo {volume} {17}},\ \bibinfo {pages} {4976} (\bibinfo {year}
  {2009})}\BibitemShut {NoStop}%
\bibitem [{\citenamefont {Kagalwala}\ \emph {et~al.}(2013)\citenamefont
  {Kagalwala}, \citenamefont {Giuseppe}, \citenamefont {Abouraddy},\ and\
  \citenamefont {Saleh}}]{Kagalwala13NP}%
  \BibitemOpen
  \bibfield  {author} {\bibinfo {author} {\bibfnamefont {K.~H.}\ \bibnamefont
  {Kagalwala}}, \bibinfo {author} {\bibfnamefont {G.~D.}\ \bibnamefont
  {Giuseppe}}, \bibinfo {author} {\bibfnamefont {A.~F.}\ \bibnamefont
  {Abouraddy}}, \ and\ \bibinfo {author} {\bibfnamefont {B.~E.~A.}\
  \bibnamefont {Saleh}},\ }\href@noop {} {\bibfield  {journal} {\bibinfo
  {journal} {Nat. Photon.}\ }\textbf {\bibinfo {volume} {7}},\ \bibinfo {pages}
  {72} (\bibinfo {year} {2013})}\BibitemShut {NoStop}%
\end{thebibliography}%

\end{document}